\documentclass[doublecol]{epl2} 

\usepackage{graphicx}
\usepackage{amsmath,amsfonts,amsthm}
\usepackage{hyperref}
\hypersetup{
    colorlinks,
    citecolor=blue,
    filecolor=blue,
    linkcolor=blue,
    urlcolor=blue
}

\title{Invariant barriers to reactive front propagation in fluid flows}

\author{F. Author\inst{1,2} \and S. Author\inst{1} \and T. Author\inst{2}}

\author{John~Mahoney\inst{1} \and Dylan~Bargteil\inst{2} \and Mark~Kingsbury\inst{2} \and Kevin~Mitchell\inst{1} \and Tom~Solomon\inst{2}}

\shortauthor{J.~Mahoney \etal}

\institute{                    
  \inst{1} School of Natural Sciences, University of California, Merced, CA 95344, USA\\
  \inst{2} Department of Physics and Astronomy, Bucknell University, Lewisburg, PA 17837, USA
}


\abstract{
We present theory and experiments on the dynamics of reaction fronts
in two-dimensional, vortex-dominated flows, for both  time-independent 
and periodically driven cases.  We find that the front propagation process is
controlled by one-sided barriers that are either fixed in the
laboratory frame (time-independent flows) or oscillate periodically 
(periodically driven flows).  We call these barriers \emph{burning invariant manifolds}
(BIMs), since their role in front propagation is analogous to
that of invariant manifolds in the transport and mixing of passive
impurities under advection.  Theoretically, the BIMs emerge from a
dynamical systems approach when the advection-reaction-diffusion
dynamics is recast as an ODE for front element dynamics.  Experimentally, we
measure the location of BIMs for several laboratory flows and confirm
their role as barriers to front propagation.
}

\begin{document}

\maketitle

Many dynamical systems are characterized by the propagation of fronts
that separate distinct phases, including chemical
reactions~\cite{tel2005}, plankton blooms~\cite{scotti07},
plasmas~\cite{beule98}, epidemics~\cite{russell04}, and flame fronts.
Fronts propagating in non-advecting {\em reaction-diffusion} (RD)
systems, \textit{i.e.}, with no fluid flow, have been the subject of
much research.  For instance, front speeds in the RD regime are well
described by the existing Fisher-Kolmogorov-Petrovsky-Piskonuv (FKPP)
theory~\cite{fisher, kpp}. However, many real systems over a broad
range of length scales exhibit coherent fluid or fluid-like motion
that dramatically impacts front propagation, \textit{e.g.}, plankton
blooms in ocean currents~\cite{Feudel08}, or chemical reactions in
microfluidic devices~\cite{Mezic07}.  Despite the importance of flows
in front-producing systems, a general framework for understanding
their effect is lacking.  Notably, attempts to extend FKPP theory
through the use of an enhanced diffusivity have been shown inadequate
in describing front propagation in laminar {\em
advection-reaction-diffusion} (ARD) systems~\cite{paoletti05b}.  
This suggests that we approach the problem from a different perspective.

In this Letter, through both theory and experiment, we reveal
fundamental geometric structures that govern front propagation in
two-dimensional (2D) flows.
We draw inspiration from the theory of chaotic
advection, which emphasizes the key role played by invariant manifolds
as barriers to passive transport~\cite{ottino89,Wiggins92}.  The
central idea of this Letter is that
analogous manifolds---what we call burning
\footnote{We use the term ``burning'' generically for any front propagation, such as the experimental chemical fronts here.} 
invariant manifolds (BIMs)---
serve as \emph{one-sided} barriers to front propagation.

\begin{figure}
\centering
\includegraphics[width=0.8\linewidth]{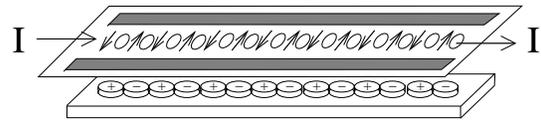}
\caption{
Exploded view of magnetohydrodynamic forcing.  Current interacts 
with alternating magnets to produce a flow composed of a chain of alternating vortices.
The fluid channel measures 1.9 cm x 27 cm.}
\label{fig:appar}
\end{figure}

We shall first consider a chain of alternating quasi-2D vortices
(fig.~\ref{fig:appar}), considering both a time-independent flow and a
time-periodic flow with vortices that oscillate laterally.  Both of
these experiments are complemented by qualitative theoretical models.
Finally we shall demonstrate the applicability of the proposed concepts
in the more general setting of a spatially disordered flow (fig.~\ref{fig:RandomFlow}).

Vortex chains provide a suitable context to introduce our geometric
approach to front propagation, since much is already known about
passive transport in these systems.  Previous studies in both
time-independent and time-periodic flows have found long-range passive
transport that is often diffusive with a variance that grows linearly
in time~\cite{solomon88, camassa91, solomon96}.  For the time-periodic
flow, passive transport has been successfully analyzed
\cite{camassa91,solomon96} in terms of invariant manifolds and the
lobes~\cite{MacKay84,Rom-Kedar90b,Wiggins92} formed by their
intersections.  Reactive front propagation in this flow has also been
experimentally studied~\cite{paoletti05b, pocheau06}.  For the
time-periodic flow, fronts often mode-lock to the external
forcing~\cite{cencini03,paoletti05b}, propagating an integer number of
vortex pairs in an integer number of drive periods.  Importantly, this
result contradicts any FKPP-type analysis that predicts front speeds
that grow monotonically with enhanced diffusivity.

The vortex chain is generated (fig.~\ref{fig:appar}) using a
magnetohydrodynamic forcing technique~\cite{solomon96}: an electric
current passing through a thin (2 mm) conducting fluid interacts with
a field produced by an alternating pattern of 1.9 cm diameter magnets
below the fluid. Two strips of plastic define the fluid channel.  The
result is a chain of 14 vortices, each with width and height of
$D=1.9$ cm.  The flow can be made time-periodic by oscillating the
magnets laterally, causing the vortices to oscillate likewise.  The
timescale of magnet oscillation is much longer than the viscous
diffusion time ($\sim$ 4 s), which is a characteristic relaxation time
for velocity fluctuations in the fluid layer.  The spatially
disordered flow (fig.~\ref{fig:RandomFlow}a) is similarly generated,
except the plastic strips are removed and the magnets are replaced by
a disordered 2D configuration of smaller (0.6 cm) magnets.

The fluid for all experiments is composed of the chemicals for the
excitable, ferroin-catalyzed, Belousov-Zhabotinsky (BZ)
reaction~\cite{boehmer08}.  The fluid is initially orange; insertion
of a silver wire triggers a green reaction
that propagates through the fluid.  The reacting fluid
is imaged from above with a CCD video camera.
The front propagation speed is $V_0 = 0.007$ cm/s in the absence of a flow.
In the theory, we assume the sharp front limit
(consistent with the experiments), meaning
the reaction proceeds rapidly compared to diffusion.
We also assume that the chemical reaction has negligible feedback on the fluid flow.

We accompany these vortex chain experiments with theoretical computations 
using the following 2D fluid velocity field~\cite{solomon88},
\begin{equation}
\begin{aligned}
u_x(x, y, t) &= + \sin(\pi[x + b \sin (\omega t)]) \cos(\pi y ), \\
u_y(x, y, t) &= - \cos(\pi[x + b \sin (\omega t)]) \sin(\pi y ),
\label{eq:velfield}
\end{aligned}
\end{equation}
where $0 \le y \le 1$, and $b \equiv B/D$, and $\omega \equiv \Omega
D/U$ are dimensionless parameters with $U$, $\Omega$, and $B$ the
(dimensional) maximum fluid speed, driving frequency, and driving
amplitude, respectively ($b = B = 0$ for a time-independent flow).
This model has free-slip boundary conditions (BCs).  While the
experimental flow has no-slip BCs, it attains velocities comparable to
the free-slip model within 1 mm of the wall.  Also, Ekman
pumping~\cite{solomon03} in our experiments
produces a weak 3D secondary
flow that is not included in the model.  Nevertheless, this model
captures the basic features of the experimental flow and, in fact, has
been successfully used to model both passive transport and
mode-locking of reaction fronts for previous experiments
~\cite{solomon88,camassa91,cencini03,paoletti05b}.
\begin{figure}
\centering
\includegraphics[width=\linewidth]{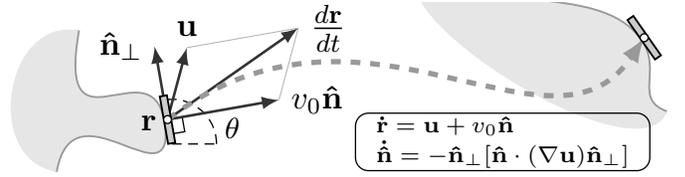}
\caption{Each reaction front element independently propagates
  forward under advection ($\mathbf{u}$) and burning ($v_0 \mathbf{\hat{n}}$). 
These 4D vector equations reduce to the 3D ODE eq.~(\ref{eq:3DODE}).}
\label{fig:3DODE}
\end{figure}
Previous theoretical studies of ARD in a vortex chain
~\cite{cencini03,abel01} utilized an Eulerian-grid-based computation.
In contrast, we directly model the dynamics of the front between
reactant and product using the following 3D
ODE~\footnote{Equation~(\ref{eq:3DODE}) can also be derived from the
  G-eqn \textit{cf.}~\cite{Oberlack10}.}
(fig.~\ref{fig:3DODE}),
\begin{subequations}
\label{eq:3DODE}
\begin{align}
\displaystyle \dot{x} &= u_x + v_0 \sin{\theta},  \quad \quad \quad
\dot{y} = u_y - v_0 \cos{\theta},
\\
\displaystyle \dot{\theta} &= -2 u_{x,x} \sin{\theta} \cos{\theta}
- u_{x,y} \sin^2{\theta}
+ u_{y,x} \cos^2{\theta},
\end{align} 
\end{subequations}
where $\mathbf{r} = (x,y)$ is the position of an infinitesimal front
element, $\theta$ is the local orientation of the front, defined with
respect to the $x$-axis, $\mathbf{u}(x,y,t)$ is the prescribed
incompressible fluid velocity field, and $v_0 \equiv V_0/U$ is the
dimensionless burning speed.  The 3D ODE can also be expressed in
vector form, as shown in fig.~\ref{fig:3DODE}.  These ODEs assume that
the front propagation speed is constant in the local fluid frame and
does not depend on the local curvature of the front \cite{Neufeld09}.
\begin{figure}
\includegraphics[width=\linewidth]{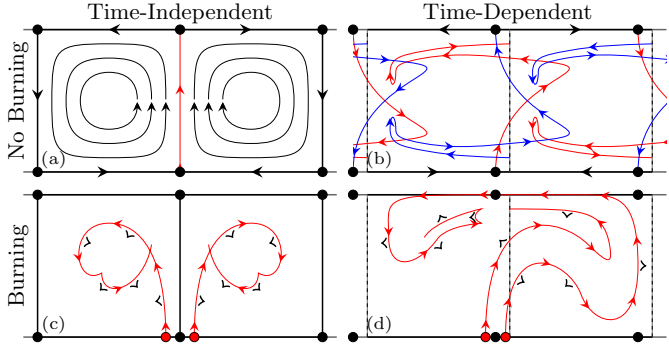}
\caption{Schematics showing the geometric structures that govern passive 
transport and front propagation. 
Black arrows indicate fluid velocity. Blue/red
denote stable/unstable manifolds. Black dots denote fixed points of
2D fluid flow.
Red dots denote burning fixed points of eq.~(\ref{eq:3DODE}).
Arrows tangent to unstable manifolds show unstable direction;
normal wedge shapes show the ``burning'' direction $\mathbf{\hat{n}}$ of BIMs.}
\label{fig:ForcingVsBurning}
\end{figure}
We investigate four physical regimes (fig.~\ref{fig:ForcingVsBurning}),
the first two of which review existing theory,
while the latter two introduce BIMs, their measurement, and their function.

\textit{Time-independent fluid flow, passive mixing}
(fig.~\ref{fig:ForcingVsBurning}a): Advection in a
regular (integrable) flow is the base case.  Here, the streamlines are
closed, forming invariant tori.  The stable and unstable invariant
manifolds, anchored to hyperbolic fixed points on the top and bottom
of the channel, are degenerate with one another and form separatrices,
dividing the channel into isolated vortex cells.

\textit{Time-periodic fluid flow, passive mixing}
(fig.~\ref{fig:ForcingVsBurning}b): Mixing in the time-periodic flow
is typically chaotic~\cite{solomon88, camassa91, solomon96};
consequently, the dynamics are now best studied by a Poincar\'{e} map
which advects a given $(x,y)$ position forward over one driving
period.  The separatrices from the time-independent case split into
separate stable and unstable invariant manifolds, each attached to one
hyperbolic fixed point on the channel wall.  Lobes formed from the
intersections of these complicated curves govern passive transport
between neighboring vortices in the flow~\cite{Wiggins92,camassa91,
  solomon96,MacKay84,Rom-Kedar90b}.

\textit{Time-independent fluid flow, reactive front propagation}
(fig.~\ref{fig:ForcingVsBurning}c): The addition of burning ($v_0 \ne
0$) results in a few critical changes, central to this Letter.  First,
each advective hyperbolic fixed point in
fig.~\ref{fig:ForcingVsBurning}a splits into two \emph{burning} fixed
points (fixed points of eq.~\ref{eq:3DODE}), one on either side.  Each
burning fixed point lives in $xy\theta$-space, and so is endowed with
a burning direction.  These occur where the fluid velocity is exactly
balanced by the burning velocity of the front.  In our model, these
points lie on the channel wall, while in our experiments, they lie
roughly 1 mm away due to the no-slip BC.  Each of these burning fixed
points has one unstable and two stable directions, generating
one-dimensional unstable manifolds -- the \emph{burning invariant
  manifolds} (BIMs) shown in fig.~\ref{fig:ForcingVsBurning}c.  It is
critical to recognize that each BIM has a burning direction, denoted
by wedge shapes.  In other words, the addition of burning splits each
manifold into a left- and right-burning BIM~\footnote{A pair of stable
  BIMs for the top, middle fixed point also exists in
  Fig.~\ref{fig:ForcingVsBurning}c, related to the unstable BIMs by
  reflection about the horizontal.  Additional stable and unstable
  BIMs exist for the other fixed points as well.}.  Note that the curves in
fig.~\ref{fig:ForcingVsBurning}c are 2D projections of BIMs in 3D,
causing the appearance of intersections and cusps.  Cusps have the
semi-cubic $y^2 = x^3$ normal form found in ray optics.

\begin{figure}
\includegraphics[width = \linewidth]{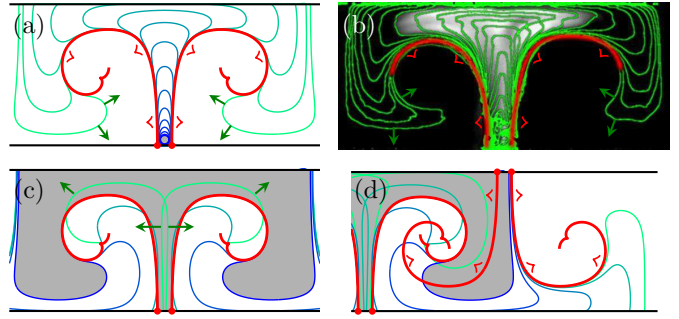}
\caption{Time-independent flow with reaction fronts ($v_0 = 0.16$). a) BIMs (red)
  calculated directly from eqs.~(\ref{eq:velfield}) and
  (\ref{eq:3DODE}).  Simulation shows reaction front evolving over
  time (blue to green), converging on the BIMs.  b)
  Experiment shows evolving reaction front (green, images 5 sec apart) and experimentally
  determined bounding BIMs (red).  Each front is extracted from a
  single reaction image, one of which is shown in white. 
  c) Simulation shows fronts
  wrapping around the BIM cusp and filling in the vortex centers.
  Oppositely oriented fronts (green) collide in the image center.  d)
  A neighboring pair of BIMs exists to the right of the original pair.
  As the front evolves right around the original BIM it encounters the
  second pair of BIMs, though it is only bounded by the BIM burning in
  the same direction as the front.  }
\label{fig:BurningNoForcing}
\end{figure}

\begin{figure}
\includegraphics[width=\linewidth]{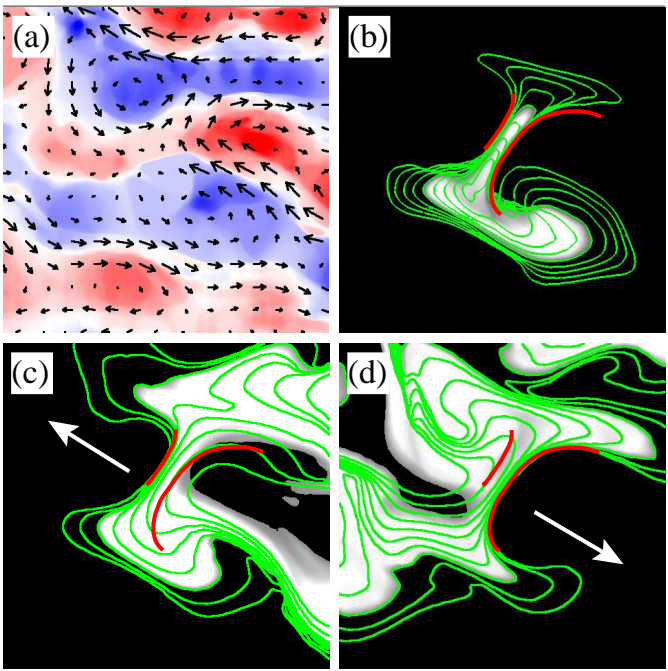}
\caption{Experiments with 3.55 cm square window of disordered flow. a) Experimental fluid
  flow measured by particle tracking. Color shows vorticity. b) As in
  fig.~\ref{fig:BurningNoForcing}b, stimulating between oppositely
  oriented BIMs produces fronts that approach and are bounded by
  BIMs. (c, d) Generic stimulations on either side burn through one BIM but not
  the other.}
\label{fig:RandomFlow}
\end{figure}

Figure~\ref{fig:BurningNoForcing}a shows simulations that illustrate
the bounding property of BIMs.  A reaction front is catalyzed at the
advective fixed point, to each side of which lies a burning fixed
point and its attached BIM.  The evolution of this front is repeatedly
plotted as it propagates away from the wall, using a computation based
on eq.~(\ref{eq:3DODE}).  Note that as the front evolves, it converges
upon the independently computed~\footnote{Numerical computation of
invariant manifolds similar to \cite{You91}.}  BIMs, with the BIMs
acting as barriers to front propagation.  The convergence
behavior is due to the fact that the BIMs are attracting in their
transverse directions.  The BIMs are \emph{one-sided} barriers,
blocking those fronts propagating in the same direction; a front
burning in the opposite direction as a BIM can pass right through, as
discussed below.  As the front reaches the projection singularity of
the BIM (the cusp), it will spiral around the singular point until the
front collides with the previously burned region, as shown in
fig.~\ref{fig:BurningNoForcing}c, thereby filling in the center of the
vortex
~\footnote{Two fronts that represent
  reactions colliding head-on have distinct $\theta$-values, and so
  have well separated trajectories under the 3D ODE
  eq.~(\ref{eq:3DODE}).  Furthermore, front element trajectories that
  reach the channel wall simply end, as the vector field is not
  defined outside the channel.}.  
Furthermore, as the front evolves to the right, into the
neighboring vortex (fig.~\ref{fig:BurningNoForcing}d),
it encounters a second pair of BIMs.
It passes through the first one (oriented opposite the front)
and is blocked by the second (aligned with the front).
Thus, BIMs are
\emph{local} barriers, since fronts can burn around a BIM segment, but
not through a BIM segment having the same burning direction.

The BIMs can be determined experimentally through a sequence of
evolving fronts (fig.~\ref{fig:BurningNoForcing}b).  These fronts are
extracted from images of a reaction, initially triggered at the bottom
fixed point.  The fronts approach a pair of curves (red), which we
identify as the experimentally measured BIMs.  Analysis of image data
confirms a drop in front speed to an order of magnitude below $V_0$ as
the front approaches the BIM, indicating that the BIMs function as
barriers.  In experiments, the BIMs are not perfect barriers due to
Ekman pumping and slight noise in the velocity field.  Experimentally,
we can not determine the BIM beyond the cusp singularity (witnessed in
the theory), since the converging front spirals around the singularity
and burns through that part of the BIM beyond the singularity.

\textit{Time-periodic fluid flow, reactive front propagation} 
(fig.~\ref{fig:ForcingVsBurning}d): As is the case for the time-independent
flow (figs.~\ref{fig:ForcingVsBurning}a, c), the addition of reactive burning to the
time-periodic flow results in the splitting of each fixed point and 
their invariant manifolds (figs.~\ref{fig:ForcingVsBurning}b, d).
The Poincar\'{e} map in fig.~\ref{fig:ForcingVsBurning}d shows left- and
right-burning fixed points, along with left- and right-burning
BIMs.

\begin{figure}
\begin{center}
\includegraphics[width=\linewidth]{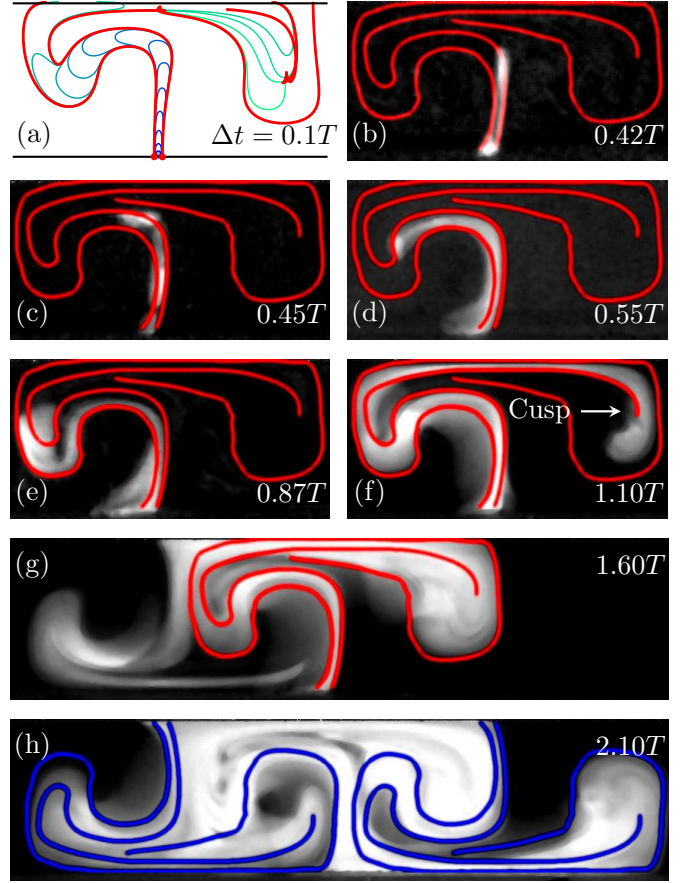}
\end{center}
\caption{Fronts in time-periodic flow.  a) Theoretical model;
  a front sequence converging onto the BIMs.  b) - h)
  Experimental images of reaction regions (white) and extracted BIMs (red).
  h) BIMs from the adjacent fixed points are shown in blue.
  For all panels, $U=0.09$ cm/s, $V_0 = 0.007$ cm/s, $B = 0.57$ cm, $\Omega = 0.16$ rad/s. 
Advanced times for trigger are given as multiples of the oscillation period $T=40$ s.}
\label{fig:ForcingAndBurning}
\end{figure}

The technique for the extraction 
of BIMs in the time-periodic case is slightly more complex than in the
time-independent case. 
Each curve in
fig.~\ref{fig:ForcingAndBurning}a is a snapshot of a simulated
evolving front, each of which was catalyzed at a different time in the
past but recorded at the common time $t=0$.  
Thus, although the initial triggering occurs at different phases of
the driving, all fronts are imaged at the same phase.  This sequence
of fronts again converges upon the BIMs (red) which act as
local barriers.

Figures~\ref{fig:ForcingAndBurning}b-\ref{fig:ForcingAndBurning}h show
an experimental realization of this protocol.  In a series of separate
experiments the reaction is triggered at different times $(t < 0)$,
and therefore different phases of the driving.  For each case, the
reaction is triggered in a region along the boundary that is mostly between
the BIMs, though since the BIMs are close together, the triggered
region sometimes overlaps the BIMs.  
Each reaction is allowed to evolve until it is imaged at $t=0$.
The red curves again show 
the experimentally extracted BIMs~\footnote{BIM
extraction from experiments: Each reaction image is smoothed,
after which a high-pass filter is applied, resulting in an
edge-enhanced image.  The sequence of edge-enhanced images is
summed, and the BIMs appear as ridges in this summed image.}.
Although eq.~(\ref{eq:velfield}) is an idealization of the
experimental flow, the BIM geometry in the model and experiment
is remarkably similar.

As seen in the time-independent flow, the BIMs form a channel
which bounds the sequence of fronts.  
Upon reaching a cusp singularity in the BIM (fig.~\ref{fig:ForcingAndBurning}f), the front sequence wraps around similar to the behavior in the time independent flow (fig.~\ref{fig:BurningNoForcing}). 
We note that in the model, the cusp is rounded in the opposite orientation (fig.~\ref{fig:ForcingAndBurning}a). 
By perturbing the model parameters, it is possible to alter this orientation.
In the next frame (fig.~\ref{fig:ForcingAndBurning}g) the reaction moves significantly left of the BIM segment shown.
We discuss the details of this mechanism below.
After burning beyond the finite BIM segments shown in fig.~\ref{fig:ForcingAndBurning}b-g, the reaction front subsequently presses against the neighboring BIMs in fig.~\ref{fig:ForcingAndBurning}h (blue) related to the red curves by the flip-shift symmetry of the vortex chain. 
Their bounding effect on the front propagation is apparent.

\begin{figure}
\begin{center}
\includegraphics[width=\linewidth]{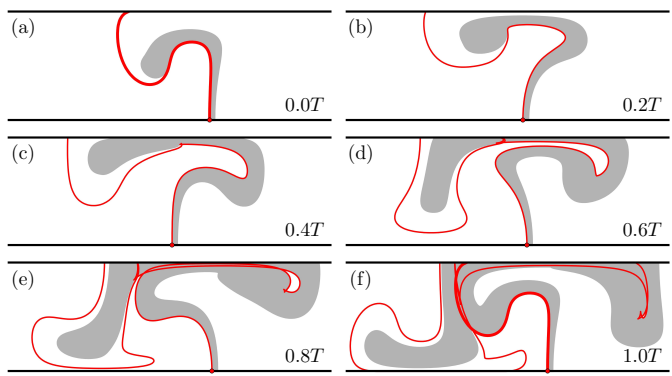}
\end{center}
\caption{a) A front (gray) begins to the right of the BIM (bold red) in a periodically driven vortex chain flow. 
A series of snapshots shows the evolution of this particular reaction over one forcing period.
The front remains bounded by the evolving BIM at each time.
Notice that in (a) the left BIM spans the channel. 
This sequence shows how the front is able to move beyond the spanning BIM.
The original BIM segment is bold in the last frame.
Note the good qualitative correspondence of fig.~\ref{fig:TimeDepBIM}a,f with fig.~\ref{fig:ForcingAndBurning}d,g.
}
\label{fig:TimeDepBIM}
\end{figure}

As the left BIM in fig.~\ref{fig:ForcingAndBurning}f spans the entire channel, it requires some additional explanation to understand how the reaction moves left of this span, since there is no cusp singularity to spiral around, as in fig.~\ref{fig:BurningNoForcing}. The simulation in fig.~\ref{fig:TimeDepBIM} demonstrates the coevolution of a single reaction front and the left BIM during the course of one complete forcing period. 
The BIM itself stretches and folds in time, generating a complicated structure that moves both to the left and right.
Upon completion of the cycle (fig.~\ref{fig:TimeDepBIM}f), the BIM maps onto itself (original segment in bold).
The reaction does not penetrate the BIM (in the burning direction) at any point during this process.
Rather, it is the extension of the BIM which allows the reaction to proceed leftward.
In fact, the left BIM not only bounds the reacted region, but also draws it along beyond the initial span, and onward down the channel. 
This process is akin to the canonical turnstile mechanism of passive transport \cite{Wiggins92}.

\begin{figure}
\includegraphics[width=\linewidth]{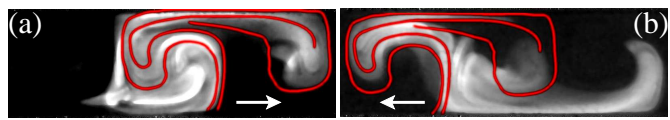}
\caption{Fronts triggered at generic points to the left a) and right b)
of a pair of BIMs.  These
images illustrate the one-sided nature of the BIMs as barriers.}
\label{fig:OneSided}
\end{figure}

As noted already for time-independent flows, the BIMs are
\emph{one-sided} barriers; reactions propagating in a direction
opposite the BIM's burning direction pass through unimpeded.
Figure~\ref{fig:OneSided}a shows a front evolving from a generic
stimulation point left of the displayed BIMs.  The front has burned to
the right, through the left-burning BIM, but is bounded by the
right-burning BIM.  Similarly, a leftward-propagating front passes
through a right-burning BIM but is blocked by the left-burning BIM
(fig.~\ref{fig:OneSided}b).

The concepts developed above are robust, since eq.~(\ref{eq:3DODE}) is
valid for any 2D incompressible flow, and BIMs are generic features of
this ODE.  We have observed the presence and influence of BIMs, both
experimentally and computationally, for a variety of parameters in the
vortex chain.  Furthermore, we have demonstrated their existence and
function using a spatially disordered, time-independent flow
(fig.~\ref{fig:RandomFlow}a).  As was illustrated for the vortex
chain, a reaction triggered near a fixed point in the disordered flow
approaches a pair of BIMs, one on either side
(fig.~\ref{fig:RandomFlow}b).  The one-sided nature of the BIMs is
also seen in figs.~\ref{fig:RandomFlow}c,d; fronts triggered outside
the displayed pair of BIMs pass through the BIM encountered first, but
stop at the second.  Other BIMs observed in this flow share these
behaviors.

Summarizing, we have
introduced {\em burning invariant manifolds} (BIMs) as geometric
objects that govern the propagation of reaction fronts in laminar
fluid flows.  We have shown that BIMs arise naturally from a
three-dimensional ODE for reaction front elements, and we have
identified BIMs in several experimental flows and have shown that they
act as one-sided barriers to front propagation.  Currently, we are
using BIMs to extend the concept of lobe
dynamics~\cite{Wiggins92,MacKay84,Rom-Kedar90b} to ARD systems.  We
are investigating the implications of BIM topology for front
propagation speeds, providing a necessary alternative to FKPP
approaches.  We are also developing a method for extracting BIMs in
time-aperiodic contexts;
this work parallels recent
studies of passive transport in which {\em Lagrangian coherent
  structures}~\cite{voth02} were used to extend invariant manifold
theory to aperiodic flows.

\acknowledgments
  These studies were supported by the US National Science Foundation
  under grants DMR-0703635, DMR-1004744, PHY-0552790, and PHY-0748828.

\bibliographystyle{eplbib}	
\bibliography{rad_bib5}  
\end{document}